\documentclass[twocolumn,pra,unsortedaddress,showpacs,floatfix,citeautoscript,nofootinbib]{revtex4-1}
\usepackage{amssymb}
\usepackage{amsmath}
\usepackage{amsbsy}
\usepackage{latexsym,epsfig,graphicx}
\usepackage{dcolumn}
\usepackage{bm}
\usepackage{graphicx}
\usepackage{subfigure}
\usepackage{color}
\usepackage{lipsum}
\usepackage[colorlinks,urlcolor=blue,citecolor=blue]{hyperref}
\usepackage{sidecap}
\sidecaptionvpos{figure}{c}

\begin{document}

\title{Quantum phases of Bose-Einstein condensates with synthetic spin--orbital-angular-momentum coupling}
\author{Chunlei Qu}
\author{Kuei Sun}
\author{Chuanwei Zhang}
\thanks{Author to whom all correspondence should be addressed: chuanwei.zhang@utdallas.edu}

\begin{abstract}
The experimental realization of emergent spin-orbit coupling
through laser-induced Raman transitions in ultracold atoms paves a
way for exploring novel superfluid physics and simulating exotic
many-body phenomena. A recent proposal with the use of
Laguerre-Gaussian lasers enables another fundamental type of
coupling between spin and orbital angular momentum (SOAM) in
ultracold atoms. We hereby study quantum phases of a realistic
Bose-Einstein condensate (BEC) with this synthetic SOAM coupling
in a disk-shaped geometry, respecting radial inhomogeneity of the
Raman coupling. We find that the experimental system naturally
resides in a strongly interacting regime in which the phase
diagram significantly deviates from the single-particle picture.
The interplay between SOAM coupling and interaction leads to rich
structures in spin-resolved position and momentum distributions,
including a stripe phase and various types of immiscible states.
Our results would provide a guide for an experimental
investigation of SOAM-coupled BECs.
\end{abstract}

\affiliation{Department of Physics, The University of Texas at
Dallas, Richardson, Texas 75080-3021, USA} \maketitle

\vspace{-0.5cm}
\section{Introduction}\label{Sec-introduction}
\vspace{-0.3cm}
The interplay between a single particle's spin and
orbital motion, or spin-orbit coupling, plays a crucial role in
various nontrivial many-body phenomena, such as topological
insulators and superconductors~\cite{Kane,Qi} in fermionic
systems, as well as exotic spinor condensates and superfluids in
bosonic ones. Considerable
effort~\cite{Lin2011,Fu2011,Pan2012,Qu2013,Ji2014,Hamner2014,Olson2014,karina2014,Campbell2015,Wang2012,Cheuk2012,Williams2013,Fu2014,Galitski2013,Zhou2013,Zhai2015,Wang2010,Wu2011,Ho2011,Zhang2012,Hu2012,Ozawa2012,Li2012,Xu2013,Zhang2013,Fetter2014,Wei2013,Gong2011,Hu2011,Yu2011,Qu13,Zhang13,Chen13,Xu2014,Lin14,Xu2014a,Jiang2014,Xu2014b}
has been devoted to the investigation of spin-orbit coupled
many-body physics in ultracold atoms due to the system's high
tunability, disorder-free environment, and, most importantly,
synthetically inducible spin-orbit coupling through the
atom--light
interaction~\cite{Spielman2009,Dalibard2011,Goldman2013}. The
recent experimental
realization~\cite{Lin2011,Fu2011,Pan2012,Qu2013,Ji2014,Hamner2014,Olson2014,karina2014,Campbell2015,Wang2012,Cheuk2012,Williams2013,Fu2014}
of spin and linear momentum (SLM) coupling with the use of
counter-propagating Gaussian lasers to induce two-photon Raman
transitions~\cite{Spielman2009} enables the direct observation and
manipulation of SLM coupled Bose-Einstein condensates (BECs) and
degenerate Fermi gases. In spin-orbit coupled BECs, interaction
effects are expected to be essential for newly emergent quantum
phases~\cite{Galitski2013}. However, compared with the typical
kinetic energy scale in the current experimental system,
\emph{i.e.}~the recoil energy $E_{L}=h^{2}/2M\lambda ^{2}$ with
atomic mass $M$ and the laser's wavelength $ \lambda \sim 1\mu$m,
the interaction effects are relatively weak such that most
behaviors of the SLM coupled BECs show little difference from the
single-particle ones~\cite{Lin2011}.

\begin{figure}[b]\vspace{-0.5cm}
\centering
\includegraphics[width=8cm]{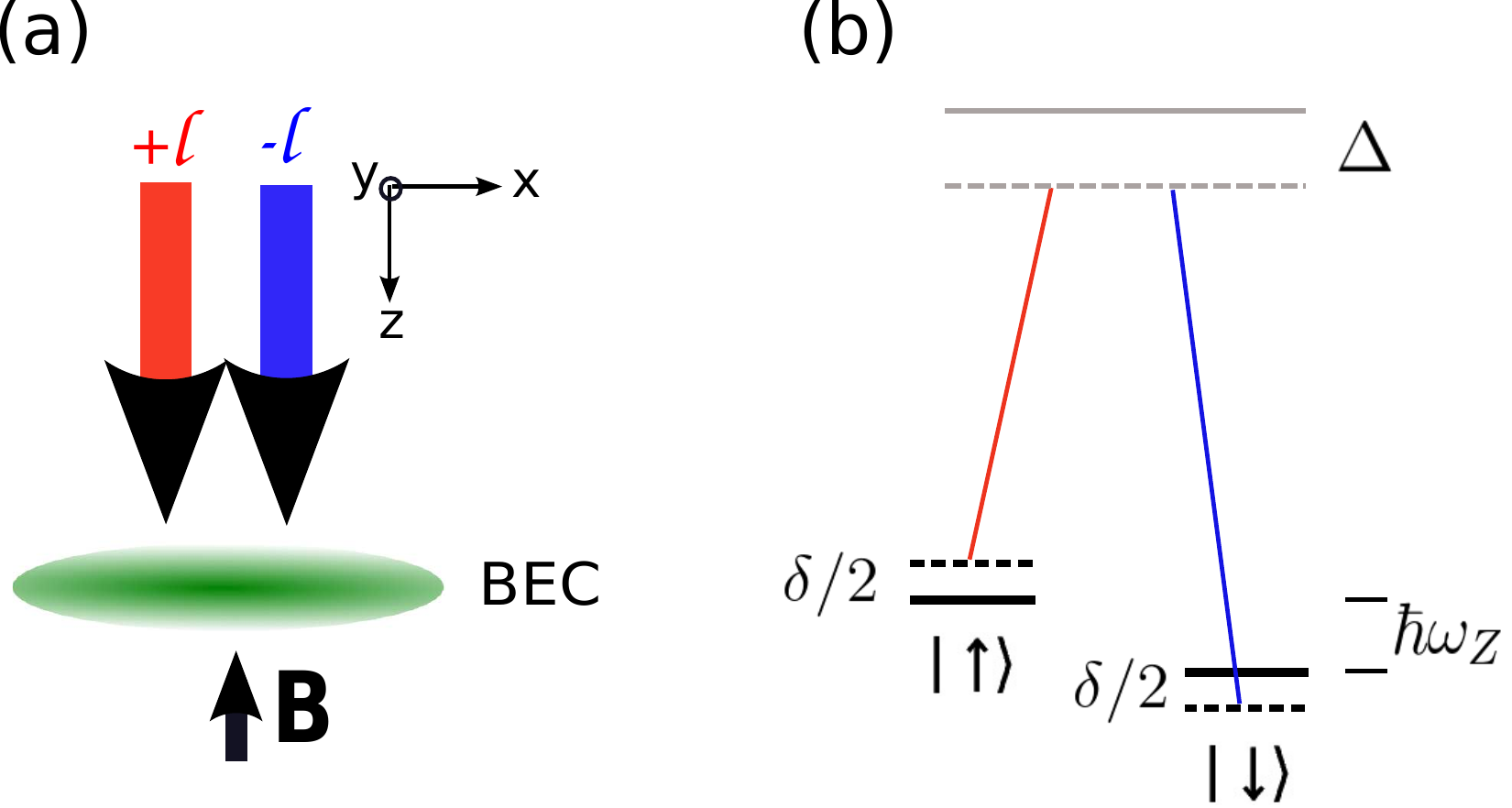}\vspace{-0.4cm}
\caption{(Color online)~(a)~Illustration for generating SOAM
coupling and (b)~Raman transitions induced by the LG laser beams.}
\vspace{-0.5cm} \label{ref-fig1}
\end{figure}

The realization of another fundamental type of spin-orbit
coupling, namely the spin--orbital angular momentum (SOAM)
coupling, in ultracold atoms has been proposed in our recent
work~\cite{Sun2014}, with the use of co-propagating higher-order
Laguerre-Gaussian (LG) lasers~\cite{Sun2014,Hu,Pu} (see
Fig.~\ref{ref-fig1}). The two LG laser beams with different
orbital angular momentum (OAM) stimulate the Raman transitions
between two atomic hyperfine states, $\left|\uparrow
\right\rangle$ and $\left|\downarrow\right\rangle$, and
concurrently impart spin-dependent OAM to the atoms
\cite{Marzlin1997,Juzeliunas2004,Cooper2010}. In such a system,
the rotational motion defines the kinetic energy scale
$E_{0}=h^{2}/2MR^{2}$ with $R\sim 15\mu$m of the order of the LG
laser's beam waist, which is much smaller than the recoil energy
$E_{L}$ and thus greatly enhances the many-body effect. Therefore,
the SOAM coupled system provides an opportunity to explore the
spin-orbit coupled physics in a strongly interacting limit. A SOAM
coupled ring BEC has been analyzed to reveal several salient
features of the aforementioned physics~\cite{Sun2014}. However,
the effect of the LG laser beams' radial distributions is
integrated out and plays no role in the ring system. This
inhomogeneous beam intensity will generate radial-dependent
potential as well as radial-dependent SOAM coupling in a
two-dimensional (2D) system, and it will result in an interesting
energetic competition against kinetic energy and interaction.

In this paper, we study the quantum phases of a realistic
$^{87}$Rb BEC in a disk-shaped geometry with SOAM coupling. We
first discuss the single-particle physics of the 2D system and
compute interacting phase diagrams as a function of Raman coupling
and detuning with experimental parameters. We then analyze the
detailed wavefunctions of each individual phase and show their
rich structures attributed to the 2D geometry and interaction. In
Sec.~\ref{Sec-model}, we build the model Hamiltonian for a SOAM
coupled BEC. The single-particle physics is discussed in
Sec.~\ref{Sec-single}. Section~\ref{Sec-BEC} presents the main
results. We also present the phase diagram of the interacting BEC
and discuss the individual phases. In Sec.~\ref{Sec-trap}, we
extend the investigation to the trapping potential and LG beams'
OAM-induced quantum phase transitions. Section~ \ref{Sec-diss}
contains the conclusion.

\vspace{-0.3cm}
\section{Model Hamiltonian}\label{Sec-model}
\vspace{-0.3cm}
We consider a pair of co-propagating LG beams of
the semiclassical form ($j=1,2$)
\begin{equation}
\Omega _{j}(\mathbf{r})=\Omega _{0}\left(
\frac{\sqrt{2}r}{w}\right) ^{\left | l_{j} \right
|}e^{-\frac{r^{2}}{w^{2}}}e^{il_{j}\phi }e^{ikz}
\end{equation}
shining on a disk-shaped BEC, where $(r,\phi ,z)$ are the three
cylindrical coordinate variables as shown in
Fig.~\ref{ref-fig1}(a). For simplicity, we assume that the two LG
beams have the same amplitudes $\Omega _{0}$ and waist $w$, but
possess opposite OAM $l_{1}=+l$ and $l_{2}=-l$. We regard the two
atomic hyperfine states as two pseudo-spins $\left|\uparrow
\right\rangle $ and $\left|\downarrow \right\rangle $,
respectively. Similar to the experiment for SLM
coupling~\cite{Lin2011}, an external Zeeman field $B$ induces a
splitting $\hbar \omega _{\rm {Z}}$ between the two states, which
also defines the frequency difference of the two LG lasers $\delta
\omega $. The two pseudo-spins are then coupled through a
two-photon Raman transition as illustrated in
Fig.~\ref{ref-fig1}b. Both LG beams are assumed to be red-detuned
from the excited state by $\Delta <0 $, while $\delta =\delta
\omega -\hbar \omega _{\rm{Z}}$ is the two-photon Raman transition
detuning.

The equilibrium and dynamic properties of the BEC with LG beams
induced SOAM
coupling are governed by the Gross-Pitaevskii (GP) equation $i\hbar \frac{%
\partial }{\partial t}\Psi =H_{\text{GP}}\Psi $, where $\Psi =(\psi
_{\uparrow },\psi _{\downarrow })^{T}$ is the spinor wave function
of the BEC. The GP Hamiltonian
\begin{equation}
H_{\text{GP}}=\frac{\mathbf{p}^{2}}{2M}+\tilde{V}_{\text{T}}(\mathbf{r})+%
\tilde{V}_{\text{LG}}(\mathbf{r})+\tilde
{V}_{\text{I}}(\mathbf{r})
\end{equation}
comprises three potential parts: (1) the external trapping potential $\tilde{V}_{%
\text{T}}$, (2) the LG laser beam potential
$\tilde{V}_{\text{LG}}$, and (3) the nonlinear mean-field
interactions $\tilde {V}_{\text{I}}$. Explicitly,
\begin{eqnarray}
&&\tilde{V}_{\text{T}}=\frac{M}{2} \left(\omega _{x}^{2}x^{2}+\omega _{y}^{2}y^{2}+\omega _{z}^{2}z^{2}\right), \\
&&\tilde{V}_{\text{LG}}=\tilde{\Omega}_{\text{R}}(r)
\begin{pmatrix}
1 & e^{-2il\phi } \\
e^{2il\phi } & 1
\end{pmatrix}
-\frac{\delta }{2}\sigma _{z}, \\
&&\tilde {V}_{\text{I}}=\left(
\begin{array}{cc}
g_{\uparrow \uparrow }|\psi _{\uparrow }|^{2}+g_{\uparrow \downarrow }|\psi
_{\downarrow }|^{2} & 0 \\
0 & g_{\uparrow \downarrow }|\psi _{\uparrow }|^{2}+g_{\downarrow
\downarrow }|\psi _{\downarrow }|^{2}
\end{array}
\right) ,
\end{eqnarray}
where $\omega _{x,y,z}$ are the trapping frequencies, and
$\tilde{\Omega}_{\text{
R}}(r)=\frac{|\Omega_1^{*}(\mathbf{r})\Omega_2(\mathbf{r})|}{4\Delta}
=\frac{\Omega _{0}^{2}}{4\Delta}\frac{2^{l}r^{2l}}{w^{2l}}\exp
\left( -\frac{2r^{2}}{w^{2}}\right) $ is the common factor of the
LG laser potential $\tilde{V}_{\rm{LG}}$, where we have included
the Stark shift potential, the Raman coupling, and the detuning
terms. The nonlinear interaction coefficients are denoted as
$g_{\sigma \sigma ^{\prime }}=4\pi \hbar ^{2}Na_{\sigma \sigma
^{\prime }}^{s}/M$, where $N$ is the total number of atoms and
$a_{\sigma \sigma ^{\prime }}^{s}$ is the $s$-wave scattering
length between the two spins $\sigma $ and $\sigma ^{\prime }$.

Similar to our previous work~\cite{Sun2014}, a unitary
transformation $\psi _{\uparrow /\downarrow }\rightarrow e^{\mp
il\phi }\psi _{\uparrow /\downarrow }$ gives
\begin{eqnarray}
{H_{{\rm{GP}}}} &\to&  - \frac{{{\hbar ^2}}}{{2M{r^2}}}{\left(
{r{\partial _r}} \right)^2} - \frac{\alpha }{{{r^2}}}{L_z}{\sigma
_z} + \frac{{\left( {L_z^2 + {\hbar ^2}{l^2}}
\right)}}{{2M{r^2}}}\nonumber\\
&&+ {{\tilde \Omega }_{\rm{R}}}(r)\left( {\begin{array}{*{20}{c}}
1&1\\
1&1
\end{array}} \right) - \frac{\delta }{2}{\sigma _z} + {{\tilde V}_{\rm{T}}}({\bf{r}}) + {{\tilde V}_{\rm{I}}}({\bf{r}}).
\end{eqnarray}
The second term in the above rotated Hamiltonian is an effective SOAM coupling of the form $\frac{\alpha }{%
r^{2}}L_{z}\sigma _{z}$, where $\alpha =\frac{\hbar l}{M}$
describes the coupling strength and $L_{z}=-i\hbar \partial _{\phi
}$ is the $z$-component OAM operator. One can see explicitly the
radial dependence $\propto r^{-2}$ of SOAM coupling, which
constitutes a key ingredient in our 2D model.

The extrema of the LG beam potential along the radial direction occurs at $R=%
\sqrt{l/2}w$. Below we use $R$, $k_{r}=1/R$, and $E_{r}={\hbar
^{2}k_{r}^{2}}/{2M}$ as the length, momentum, and energy units,
respectively. The LG beam potential can thus be written in a
dimensionless form
\begin{equation}
V_{\text{LG}}=\Omega e^{-lr^{2}}r^{2l}\left(
\begin{array}{cc}
1 & e^{-2il\phi } \\
e^{2il\phi } & 1
\end{array}
\right) ,
\end{equation}
where $\Omega =\frac{\Omega _{0}^{2}l^{l}}{4\Delta E_{r}}$ is the
only dimensionless variable for tuning $V_{\text{LG}}$ at fixed
$l$. In addition, we consider a harmonic trap $\omega _{x}=\omega
_{y}\ll \omega _{z}$ that strongly confines the BEC along the $z$
direction, so the system is well described by a 2D GP Hamiltonian.
The 2D trapping potential can be written in a dimensionless form
as $V_{\rm{trap}}(r)=\frac{1}{2}Kr^{2}$, where $K=M^{2}\omega
_{x}^{2}l^{2}w^{4}/2\hbar ^{2}$ is the dimensionless trapping
strength. In the next section, we will study the single-particle
physics with the dimensionless expression and present the ground
state diagram as a function of LG beam as well as the external
trapping potentials.

\vspace{-0.3cm}
\section{Non-interacting case}
\label{Sec-single} \vspace{-0.3cm}

\begin{figure}[t]
\vspace{-0.4cm} \centering
\includegraphics[width=8.6cm]{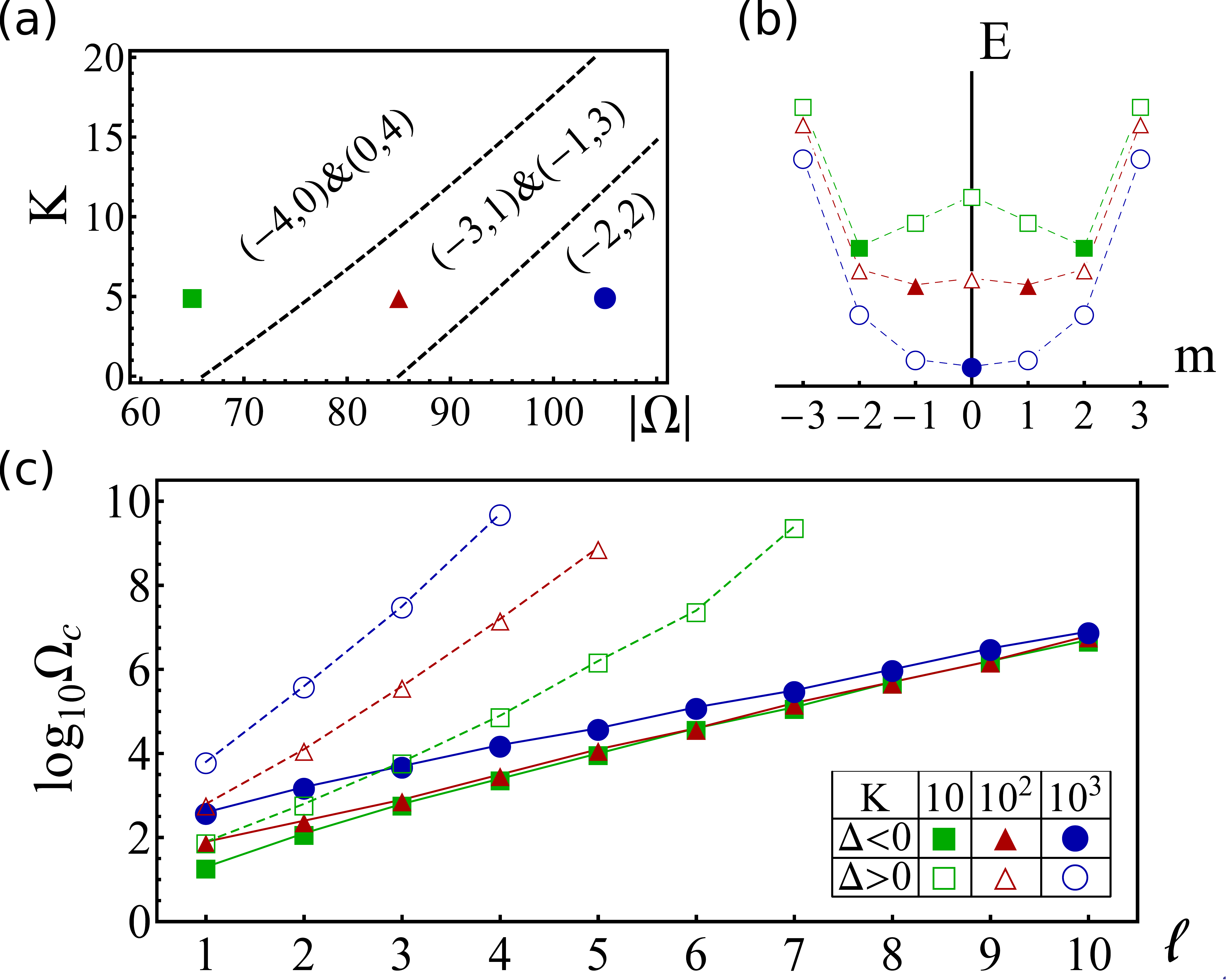}\vspace{-0.4cm}
\caption{(Color online) (a) Non-interacting state diagram in
$K$-$|\Omega|$ plane. Different states denoted with OAM quantum
numbers $(m_\uparrow,m_\downarrow)$ separate by dashed lines. (b)
Energy spectrum of the lowest band $n=0$ as a function of
quasi-OAM $m$ at select points (square, triangle, and circle) in
(a). The filled symbols denote the ground states. (c) Critical
value of
Raman coupling from double-minima to single-minimum transition for blue ($%
\Delta>0$) and red ($<0$) detuned lasers and for different
trapping strengths $K$.} \vspace{-0.4cm} \label{ref-fig2}
\end{figure}

Without interactions, $\tilde V_{\text{I}}=0$, and the GP
Hamiltonian becomes a Schr\"{o}dinger one. Due to cylindrical
symmetry, the eigenstates can be described by an angular quantum
number $m$ as well as a radial one $n$ and can be written in a
form of ${\left( {
\begin{array}{cc}
{{\chi _{\uparrow ,{n}}}(r){e^{i{m_{\uparrow }}\phi }}} , {{\chi
_{\downarrow ,{n}}}(r){e^{i{m_{\downarrow }}\phi }}}
\end{array}
}\right) ^{T}}$ with $m_{\uparrow /\downarrow }=m\mp l$~\cite{Pu}.
The angular number $m$ is an integer because OAM quantization of
each spin component requires $m_{\uparrow /\downarrow }$ to both
be integers. The OAM difference $m_{\uparrow }-m_{\downarrow
}=-2l$ comes from the effect of the SOAM coupling. The radial
number $n$ is a non-negative integer due to the radial confinement
and describes the number of radial nodes of the wave function. For
a given $m$, the radial part of the wavefunctions $\Psi
_{n}(r)={\left( {
\begin{array}{cc}
{{\chi _{\uparrow ,{n}}}} & {{\chi _{\downarrow ,{n}}}}
\end{array}
}\right) ^{T}}$ and corresponding eigenenergies $E_{n,m}$ can be obtained by
solving a radial Schr\"{o}dinger equation,
\begin{eqnarray}
E\Psi  &=&\Big[-\frac{1}{r}{\partial _{r}}\left( {r{\partial _{r}}}\right) +%
\frac{K}{2}{r^{2}}+\frac{1}{{{r^{2}}}}\left( {
\begin{array}{cc}
{m_{\uparrow }^{2}} & 0
\\
0 & {m_{\downarrow }^{2}}
\end{array}
}\right)   \nonumber \\
&&+\Omega {e^{-l{r^{2}}}}{r^{2l}}\left( {
\begin{array}{cc}
1 & 1  \\
1 & 1
\end{array}
}\right) \Big]\Psi ,  \label{eq:non-int_H}
\end{eqnarray}
where we assume $\delta =0$. We find that $E_{n,m}$ always increases with $n$%
, so it is convenient to interpret the energy spectrum by regarding $n$ as a
band index. The ground state thus lies in the lowest band of $n=0$. If we
treat $m$ as a quasi-OAM number, the lowest band structure $E_{0,m}$ can
exhibit either double minima at $m=\pm m_{0}$ with $0<m_{0}\leq l$ or a
single minimum at $m=0$, determined by $\Omega $ and $K$.

In Fig.~\ref{ref-fig2}(a), we plot the ground state diagram in
$|\Omega |$-$K$ plane for the case of $l=2$ and $\Delta <0$. Each
region is labeled by the ground-state OAM numbers $(m_{\uparrow
},m_{\downarrow })$ for each spin species. At a fixed $K$, the
system undergoes transitions from the degenerate ground states
$m=\pm 2$ to $m=\pm 1$, and to the single one $m=0$ as $|\Omega |$
increases. We also find that the spin polarization has the same
sign as $m$ and becomes zero if $m=0$. Both characteristics show
similar physics to that in SLM coupled BEC or SOAM coupled ring
BEC. Figure \ref{ref-fig2}(b) shows the lowest band for selected
points at $K=5$ in Fig.~\ref{ref-fig2}(a). One can see how the
band structure evolves between different ground states. At a fixed
$\Omega $, we find that $K$ competes with $\Omega $ and moves the
system from the single-minimum region toward the double-minimum
ones. This trend can be understood as a result of energetic
competition. The external trap tends to concentrate the BEC around
the central region to reduce the potential energy, but rotational
energy from the centrifugal barrier $r^{-2}m_{\uparrow /\downarrow
}$ in Eq.~(\ref{eq:non-int_H}) strongly increases unless the OAM
number vanishes. As a result, the system favors the majority spin
component having a low or zero angular momentum and localizing
around the trap center, even with a certain energetic penalty from
the minority component and from SOAM coupling. In
Fig.~\ref{ref-fig2}(c), we plot the critical strength $\Omega
_{c}$ of SOAM coupling for the transition between degenerate and
non-degenerate ground states as a function of $l$ at various $K$
and different signs of $\Delta $. We see that $\Omega _{c}$
increases with increasing $l$, increasing $K$, or blue-detuned
$\Delta >0$, consistent with the same physical picture of BEC's
avoiding higher rotational energy. Note that the blue-detuned
$\Delta $ centers the BEC more because the Stark shift creates a
ring-shaped barrier around the high intensity region of the LG
beams.

Before studying the interaction effect, we comment on an
interesting case, where the difference between two LG beams'
azimuthal indices $l_{1}-l_{2}$ is an odd number. In this case,
$m_{\uparrow /\downarrow }=m\mp (l_{1}-l_{2})/2$ indicates that
$m$ should be a half-integer to satisfy OAM quantization. As
$\Omega $ increases from zero, the magnitude of $m$ suppresses
from $|(l_{1}-l_{2})/2|$ until $m=\pm 1/2$. Therefore, the lowest
band always has a double-minima structure and the system always
has two degenerate ground states with opposite spin polarization.

\vspace{-0.3cm}
\section{Interacting case}
\label{Sec-BEC}
\vspace{-0.3cm}

\subsection{Phase diagram}\vspace{-0.3cm}

In this section, we focus on a realistic $^{87}$Rb system, compute
the full phase diagram, and characterize each individual phase as
well as various phase transitions. We consider two atomic internal
states $|\uparrow \rangle =|F=1,m_{F}=0\rangle $ and $|\downarrow
\rangle =|F=1,m_{F}=-1\rangle $ of $^{87}$Rb atoms that are
coupled through the Raman transitions induced by the two LG laser
beams with $l=2$. The radial intensity waist of the LG beam is
$w=12\sqrt{2}\mu m$, which gives $R\approx 17\mu m$ and
$E_{r}=2\pi \hbar \times 0.2$ ${\text{Hz}}$. The $z$-direction
trapping frequency $\omega _{z}=2\pi \times 600\text{Hz}$ is set
large enough such that the system has an effective 2D
geometry~\cite{Ramanathan2011}, while the transverse trapping
frequency $\omega _{x}=\omega _{y}=1.5\sim 6.0\text{ Hz}$ is set
isotropic. The $s$-wave scattering lengths are
$a_{\uparrow\uparrow}= 100.86 a_{B}$ and
$a_{\downarrow\downarrow}=a_{\uparrow\downarrow}= 100.4 a_{B}$ for
$^{87}$Rb, where $a_{\text{B}}$ is the Bohr radius. In a case of
particle number $N=10^{4}$, the reduced 2D nonlinear interaction
energy $E_{\text{I}}\gg E_{r}$ is much larger than the
single-particle rotational energy. Such strong nonlinear
interaction will drastically modify the single-particle physics,
as we will see. Note that with this setup, the three nonlinear
coefficients satisfy $g_{\uparrow \uparrow }g_{\downarrow
\downarrow }>g_{\uparrow \downarrow }^{2}$, which indicates that a
rotating spinor BEC without the Raman dressing should be
miscible~\cite{Ho1996}. However, we shall see that the
wavefunctions of opposite spins can be spatially separated or
immiscible in the presence of SOAM coupling.

\begin{figure}[t]
\vspace{-0.4cm} \centering
\includegraphics[width=8.6cm]{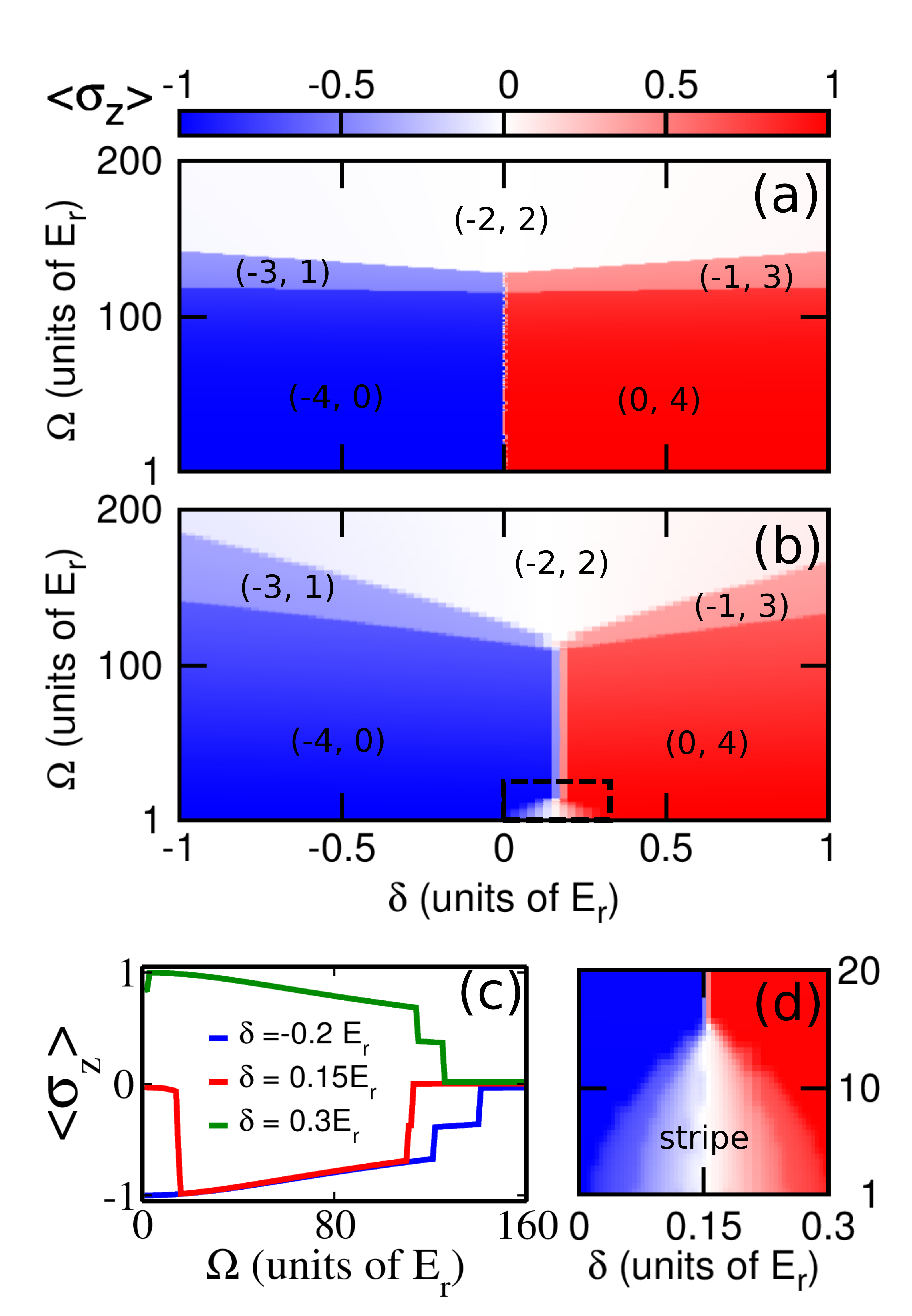}\vspace{-0.5cm}
\caption{(Color online) Phase diagram of the SOAM coupled BEC in
$\Omega$--$\delta$ plane for (a) non-interacting system and (b)
realistic BEC. The colors defined in the bar graph represent spin
polarization $\langle \sigma_z \rangle$. A stripe phase with small
polarization occurs at the bottom of (b). The other phases are
labled by their OAM numbers of the two spin components
$(m_{\uparrow },m_{\downarrow })$. (c) Spin polarization as a
function of Raman coupling $\Omega $ for detuning $\protect\delta %
=-0.2$ ,$0.15$, and $0.3E_{r}$. (d) Zoom-in of the dashed region
in (b) showing the stripe phase. Other parameters
are $\protect\omega _{x}=\protect\omega _{y}=1.5$Hz, $N=1\times 10^{4}$, and $%
l=2$. } \vspace{-0.8cm}\label{ref_fig3}
\end{figure}

\begin{figure}[t]
\vspace{-0.2cm} \centering
\includegraphics[width=8.6cm]{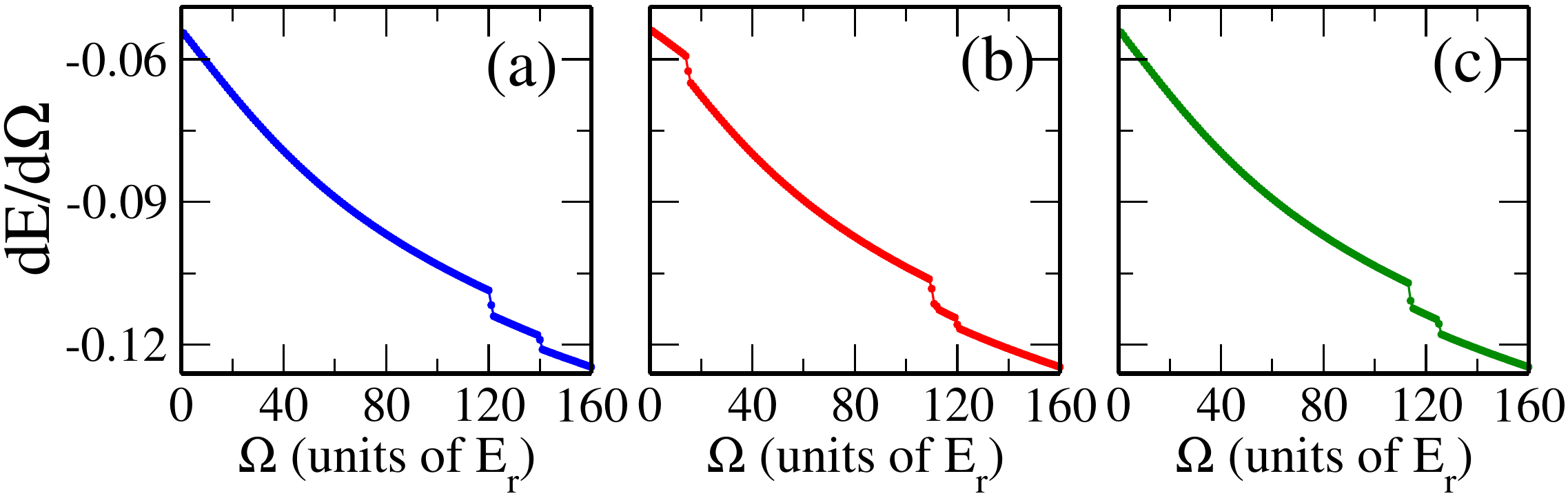}\vspace{-0.3cm}
\caption{(Color online) The derivative of ground state energy
$dE/d\Omega$ as a function of Raman coupling strength $\Omega$ for
(a) $\delta=-0.2E_r$, (b) $\delta=0.15E_r$, (c) $\delta=0.3E_r$.
Compared with Fig.~\ref{ref_fig3}(c), we see the phase transition
is first order where the energy derivative is discontinuous
whenever the spin polarization jumps.}\vspace{-0.4cm}
\label{ref_QPT}
\end{figure}

We obtain the ground state by solving the GP equation using the
imaginary-time evolution. Figure~\ref{ref_fig3} shows the phase
diagram in the Raman coupling and detuning $\Omega$-$\delta $
plane, where the colors denote the spin polarization. For
comparison, we present the non-interacting and interacting phase
diagrams in Figs.~\ref{ref_fig3}(a) and \ref{ref_fig3}(b),
respectively. We see that there are four different phases in the
interacting case according to the spin polarization $\langle
\sigma _{z}\rangle $ and OAM of the two spins $(m_{\uparrow
},m_{\downarrow })$: (i) A stripe phase with a density modulation
in the angular direction that is the superposition of $(-4,0)$ and
$(0,4)$ states; (ii) a spin-polarized phase, either $(-4,0)$ or
$(0,+4)$; (iii) a spin-polarized phase, either $(-3,+1)$ or
$(-1,+3)$; and (iv) a spin-balanced phase $(-2,2)$. We comment
that the energy difference between these phases can be small
(compared with typical trapping potential energy, SOAM coupling
energy, and kinetic energy), and BEC may end up with a metastable
state in our numerical simulation. To obtain the ground-state
phase boundaries accurately, we start from several initial states
to do the imaginary-time evolution and choose a final state with
the lowest energy as the true ground state of the interacting BEC.
The convergence tolerance of the GP calculation is chosen as
$10^{-4}E_r$ such that the energy difference between the ground
state and any metastable state is clearly discernible. We also
emphasize that the ground state of the system is robust against
small perturbations due to both the energy gap from other
metastable states and OAM quantization. Compared with the
single-particle phase diagram of Fig.~\ref{ref_fig3}(a), we see
that the interaction shifts the phase boundaries and induces a
stripe phase [Fig.~\ref{ref_fig3}(b)] that is described by a state
with density modulations. Different from the SLM coupling case,
the density modulation here is along the angular direction. In
Fig.~\ref{ref_fig3}(c), we plot the spin polarization as a
function of Raman coupling for various detunings.
Figure~\ref{ref_fig3}(d) is the enlarged stripe phase where the
spin polarization is zero or small. The reason why the stripe
phase is not symmetric around $\delta =0$ is because of the
scattering length difference of the two spins, which contributes
an effective detuning, similar to the SOAM coupled ring
BEC~\cite{Sun2014}. Here we see the premium detuning value is
about $\delta =+0.15E_{r}$. In Fig.~\ref{ref_QPT}, we plot the
energy derivative $dE/d\Omega$ as a function of Raman coupling.
Compared with Fig.~\ref{ref_fig3}(c), we see that the energy
derivative is discontinuous whenever the spin polarization jumps
or OAM changes, indicating that the phase transitions are first
order.

\begin{figure*}[t!]
\vspace{-0.4cm}
 \centering
\includegraphics[width=14cm]{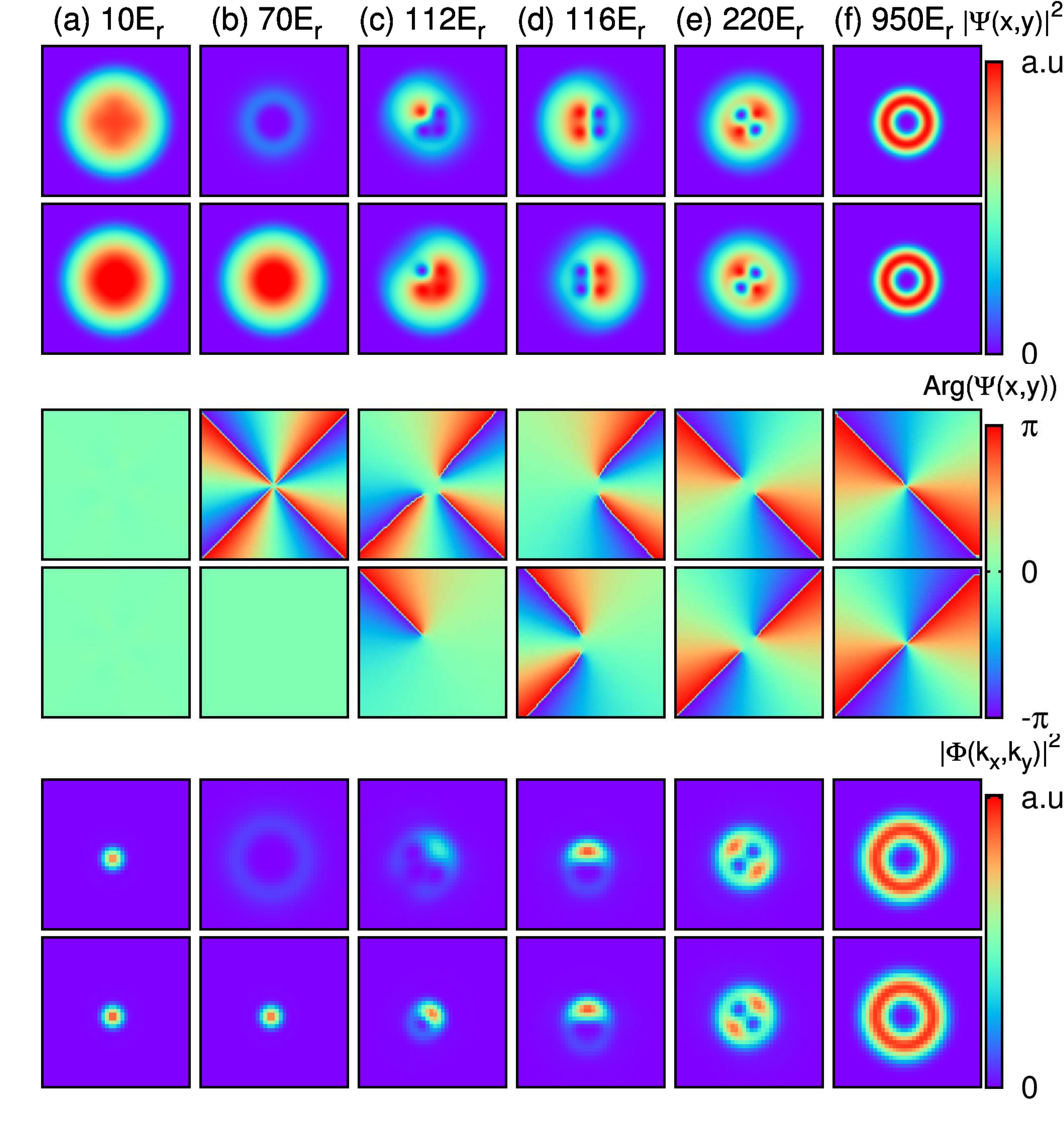}\vspace{-0.7cm}
 \caption{(Color online) Spin-resolved density
distributions of up (upper panel) and down (lower) spins at Raman
coupling strengths (a) $10E_{r}$, (b) $70E_{r}$, (c) $112E_{r}$,
(d) $116E_{r}$, (e) $220E_{r}$, and (f) $950E_{r}$. The top two
rows show the real space density distributions, the middle two do
the wavefunction phases, and the bottom two do the momentum
distributions. The Raman detuning is $\delta =+0.15E_{r}$. The
displayed range of the axis is $[-57\mu m, +57\mu m]$ in real
space and $[-6k_r, 6k_r]$ in momentum space. All other parameters
are the same as Fig.~\ref{ref_fig3}.} \vspace{-0.4cm}
\label{xspace}
\end{figure*}

We shall see in the next section that each phase can exhibit
richer structures in the spin-density distribution, including a
phase separation in a large parameter region: the two spins can be
immiscible due to SOAM coupling and interaction. In this case, the
wave function of the two spins is not exactly of the form $\chi
_{\sigma }(r)e^{im_{\sigma }\phi }$ as in the non-interacting
case. However, one can still find a set of integer winding numbers
$(m_{\uparrow },m_{\downarrow })$ by integrating the phase of each
spin's wavefunction over a path enclosing the system. Such winding
numbers satisfy $m_{\uparrow }-m_{\downarrow }=-2l$ and lock with
the spin polarization. Therefore, similar to the OAM numbers in
the non-interacting case, they are good quantum numbers for
distinguishing different phases in the presence of interaction,
although the system may show different types of phase separation
at a given $(m_{\uparrow },m_{\downarrow })$.

\vspace{-0.3cm}
\subsection{Individual phases}
\vspace{-0.3cm}

In this section, we discuss the individual phases along the
$\delta=+0.15E_r$ line in the phase diagram of
Fig.~\ref{ref_fig3}(b) in more details.

For very week Raman coupling and small detuning, we see that the
system prefers a stripe phase as shown in Fig.~\ref{xspace}(a).
The two spins have density modulations in the angular direction,
which arise from the fact that there are two degenerate dressed
states with similar weight at quasi-OAM $m=-2$ and $m=+2$ [green
squares in Fig.~\ref{ref-fig2}(b)] that interfere with each other
in the real space. Because of the superposition of two dressed
states, the phase [Arg($\Psi $)] of the two spins in the stripe
phase is almost uniform and their momentum space distributions
($|\Phi |^{2}$) are centered at $\mathbf{k}=0$.

With the increase of the Raman coupling, there is a competition
between the stripe phase and the occupation of a single
dressed-state phase. In the SLM coupling case, the stripe phase
occupies a small parameter region and the transition to the single
dressed state occurs around $\Omega \sim 0.2E_{r}$. However, here
we see that the critical value for this transition is about
$\Omega \sim 15E_{r}$ in the SOAM coupling system. When the
particle number or the angular momentum $l$ is increased, this
parameter region may be even larger. Note that the stripe phase
has never been observed experimentally yet in the SLM coupling
system because of its small parameter region and the stringent
requirement to resolve the density modulations separations (which
are at the order of $\lambda \sim 1\mu m$). It might be easier to
resolve the density modulation separations (which are at the order
of $R\sim 10\mu m$) in the SOAM coupled BEC.

\begin{figure*}[t!]\vspace{-0.4cm}
\centering
\includegraphics[width=14cm]{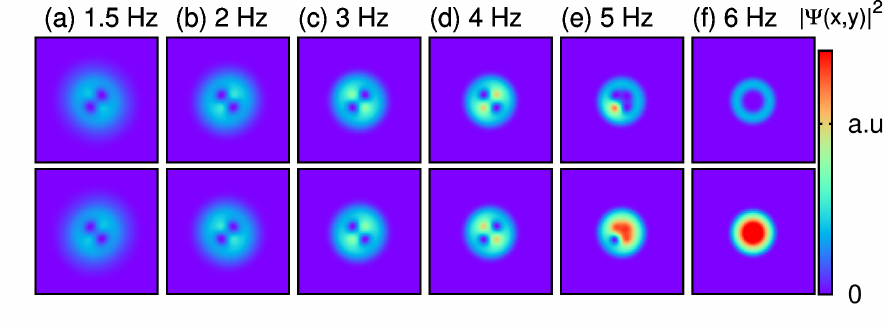}\vspace{-0.7cm}
\caption{(Color online) Density distribution of spin up (upper
panel) and spin down (lower) for different trapping frequencies
(a) $1.5\text{Hz}$,
(b) $2\text{Hz}$, (c) $3\text{Hz}$, (d) $4\text{Hz}$, (e) $5\text{Hz}$, and (f) $6%
\text{Hz}$ at $\protect\delta =0E_{r}$ and $\Omega =250E_{r}$. The
displayed range of the axis is $[-57\mu m, +57\mu m]$ in real
space. All other parameters are the same as in
Fig.~\ref{ref_fig3}.} \vspace{-0.0cm} \label{fig-trap}
\end{figure*}

\begin{figure*}[t!]\vspace{-0.2cm}
\centering
\includegraphics[width=14cm]{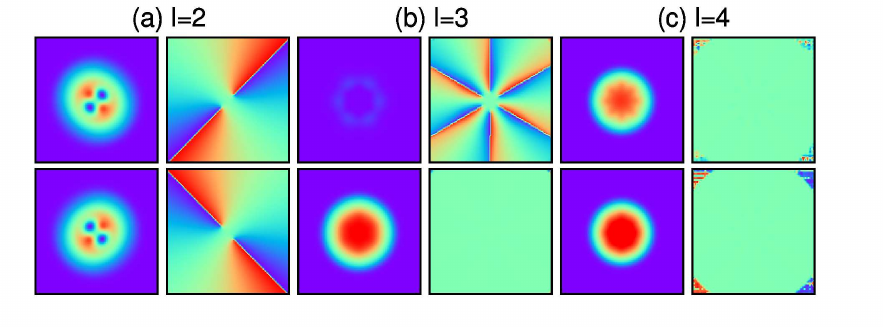}\vspace{-0.7cm}
\caption{(Color online) Real-space density distribution and phase
of spin up (upper panel) and spin down (lower panel) for LG lasers
of different OAM (a) $l=2$, (b) $l=3$, and (c) $l=4$ at $\delta
=0$ and $\Omega =250E_{r}$. For each figure, the left side is the
real-space density distribution and the right side is the phase of
the wave function. The displayed range of the axis is $[-57\mu m,
+57\mu m]$ in real space. All other parameters are the same as in
Fig.~\ref{ref_fig3}.}\vspace{-0.4cm} \label{fig-OAM}
\end{figure*}

In Fig.~\ref{xspace}(b), we show a single dressed-state phase,
where one of the spin states does not carry any OAM and the other
spin carries the largest OAM, \emph{i.e.}~$m_{\uparrow }=0$ and
$m_{\downarrow }=4$. Because of their different OAM, the density
of up spin is of a rotationally symmetric Gaussian shape, while
that of down spin is of a ring shape with a hollowed center.
Similar features are obtained in the momentum space density
distributions.

When the Raman coupling is strong enough, the two spins may both
carry OAM as shown in Figs.~\ref{xspace}(c)--\ref{xspace}(e),
where we have $(m_{\uparrow },m_{\downarrow })=(-1,+3)$ or
$(-2,+2)$. This transition can be understood from the
single-particle physics picture, where the lowest-energy band
evolves from a double minima region to a single minimum region.
However, an interesting result shows that the two spin components
appear to be immiscible or spatially separated, \emph{i.e.},
wherever one spin dominates, the other spin is strongly
suppressed. The phase separation occurs in a large parameter
region of $\Omega = 110 \sim 800 E_{r}$ in Fig.~\ref{ref_fig3}(b)
when both spin components carry OAM. The emergence of phase
separation can
be understood from the perspective of lowering nonlinear interaction energy $%
\sim \int \int (\rho _{\uparrow }+\rho _{\downarrow })^{2}dxdy$.
When the two spins both carry angular momentum and are not
separated, the total density of the system must have a hollowed
density at the trap center. The total nonlinear interaction energy
of this system is usually larger than the separated phase where
the total density distribution does not have a hollowed center but
instead is in a Gaussian shape~\cite{Ho1996,Ueda}.

Among these immiscible states, the two spins carrying opposite OAM
$+2$ and $-2$ display very interesting density distributions, as
shown in Figs.~\ref{xspace}(d) and \ref{xspace}(e): the two holes
and two peaks of the two spin components could appear in a dipole
(d) or quadrupole (e) form. We also verify that there are more
complicated and rich separated phases for larger $l$. The momentum
distributions of each of the two states are quite different and
may be distinguished in the time-of-flight experiments.

In Fig.~\ref{xspace}(f), we plot the density distributions of the
two spins for a very large Raman coupling $\Omega =950E_{r}$. When
the LG beam intensity is too strong, $\Omega > 800 E_r$, both
components are trapped around the potential minimum of
$V_{\rm{LG}}$ and forced to mix with each other. Note that the
Raman coupling and the Stark shift form an effective ring-shaped
optical trap, which is even deeper than the harmonic trap center
for such a larger Raman coupling strength. In this case, the two
components exhibit a miscible ring distribution and their momentum
distributions are also of a ring-shape, which returns to the ring
limit~\cite{Sun2014}.

\vspace{-0.3cm}
\section{Harmonic trap and OAM induced phase transition}
\label{Sec-trap} \vspace{-0.3cm}

In this section, we study the effects of the harmonic trapping
potential and larger OAM number of the LG beams. From the
single-particle results in Fig.~\ref{ref-fig2}(c), we see that the
trapping strength may drive a quantum phase transition between the
states with different OAM. In Fig.~\ref{fig-trap}, we plot the
real-space density distributions of the two spins with increasing
trapping potential from $1.5$Hz to $6$Hz. The two spins are pushed
inward when the trapping potential is increased and the OAM of the
two spins changes from $(-2,2)$ to $(-3,1)$ and then $(-4,0)$,
which is consistent with the energetic competition described in
Sec.~\ref{Sec-single}. In previous SLM coupled systems, the
trapping strength was relatively weak and thus there were no
observable effects with the change of trapping frequencies. Our
results demonstrate that the SOAM coupled BEC has a significant
response to the trapping potential.

In Fig.~\ref{fig-OAM}, we plot the density distributions and
phases of the two spins for (a) $l=2$, (b) $l=3$, and (c) $l=4$.
We see that the two spins are in a $(-2, 2)$ separated phase for
$l=2$, then in the $(-6, 0)$ spin-polarized miscible phases for
$l=3$, and eventually a stripe phase from the superposition of
$(-4, 0)$ and $(0, +4)$ dressed states for $l=4$. This can be
understood from the $l$ dependence of the critical Raman coupling
strength, similar to the single-particle results in
Fig.~\ref{ref-fig2}(c).

\vspace{-0.3cm}
\section{Conclusion}
\label{Sec-diss} \vspace{-0.3cm}

In conclusion, we studied the quantum phases of a SOAM coupled BEC
with realistic experimental parameters in a 2D geometry. The phase
diagram shows distinct physics from a SLM coupled BEC due to the
naturally strong interaction and OAM quantization, and even from a
SOAM coupled ring BEC due to the radial inhomogeneity. The
interaction-induced stripe phase and different immiscible states
have been characterized by the spin-resolved position and momentum
distributions. Various quantum phase transitions induced by the
Raman coupling, detuning, trapping potential, or LG beams' OAM
have been reported too. Although the results are presented for
$^{87}$Rb atoms, our model and treatment can be easily applied on
others such as $^{23}$Na. For a typical Raman coupling strength of
1 kHz and an experimental time scale of 1 s, the heating effects
in both $^{87}$Rb and $^{23}$Na systems should be
weak~\cite{Wei2013,Sun2014}.

Since the dynamics of ultracold atoms under the application of LG
beams have been experimentally investigated by several
groups~\cite
{Andersen2006,Ryu2007,Leslie2009,Gullo2010,Beattle2013,Moulder2012},
and ongoing experimental effort is along this direction for the
realization of SOAM coupling~\cite{DAMOP}, the achievement of a
BEC ground state with SOAM coupling would be expectable in the
near future. Our results hence provide timely predictions for
experimental observation.

\textbf{Acknowledgements}: We are grateful to L. Jiang and Y. Xu
for interesting discussions. This work is supported by ARO
(W911NF-12-1-0334) and AFOSR (FA9550-13-1-0045).


\vspace{-0.5cm}
\begin{thebibliography}{99}
% Topological insulators and superconductors

\bibitem{Kane} M. Z. Hasan and C. L. Kane, \emph{Colloquium: Topological insulators}, \href{http://dx.doi.org/10.1103/RevModPhys.82.3045}%
{Rev. Mod. Phys. \textbf{82}, 3045 (2010)}.

\bibitem{Qi} X.-L. Qi and S.-C. Zhang, \emph{Topological insulators and superconductors},\href{http://dx.doi.org/10.1103/RevModPhys.83.1057}%
{Rev. Mod. Phys. \textbf{83}, 1057 (2011)}.

% SOC experiments on bosons

\bibitem{Lin2011} Y.-J. Lin, K. Jim\'{e}nez-Garc\'{\i}a and I. B. Spielman, \emph{Spin-orbit-coupled Bose-Einstein condensates},
\href{http://dx.doi.org/10.1038/nature09887}{Nature, \textbf{471}, 83-86
(2011)}.

\bibitem{Fu2011} Z. Fu, P. Wang, S. Chai, L. Huang, and J. Zhang, \emph{Bose-Einstein condensate in a light-induced vector gauge potential using 1064-nm optical-dipole-trap lasers}, \href{http://dx.doi.org/10.1103/PhysRevA.84.043609}%
{Phys. Rev. A \textbf{84}, 043609 (2011)}.

\bibitem{Pan2012} J. -Y. Zhang, S.-C. Ji, Z. Chen, L. Zhang, Z. -D. Du, B.
Yan, G.-S. Pan, B. Zhao, Y. -J. Deng, H. Zhai, S. Chen, and J. -W. Pan, \emph{Collective Dipole Oscillations of a Spin-Orbit Coupled Bose-Einstein Condensate},
\href{http://dx.doi.org/10.1103/PhysRevLett.109.115301}{Phys. Rev. Lett.
\textbf{109}, 115301 (2012)}.

\bibitem{Qu2013} C. Qu, C. Hamner, M. Gong, C. Zhang, and P. Engels, \emph{Observation of Zitterbewegung in a spin-orbit-coupled Bose-Einstein condensate}, \href{http://dx.doi.org/10.1103/PhysRevA.88.021604}%
{Phys. Rev. A \textbf{88}, 021604(R) (2013)}.

\bibitem{Ji2014}S.-C. Ji, J.-Y. Zhang, L. Zhang, Z.-D. Du, W. Zheng, Y.-J. Deng, H. Zhai, S. Chen, and J.-W.
Pan, \emph{Experimental determination of the finite-temperature
phase diagram of a spin-orbit coupled Bose gas},
\href{http://dx.doi.org/10.1038/nphys2905}{Nat. Phys. \textbf{10},
314 (2014)}.

\bibitem{Hamner2014} C. Hamner, C. Qu, Y. Zhang, J. Chang, M. Gong, C.
Zhang, and P. Engels, \emph{Dicke-type phase transition in a
spin-orbit-coupled Bose Einstein condensate}, \href{http://dx.doi.org/10.1038/ncomms5023}{Nat.
Commun. \textbf{5}, 4023 (2014)}.

\bibitem{Olson2014} A. J. Olson, S.-J. Wang, R. J. Niffenegger, C.-H. Li, C.
H. Greene, Y. P. Chen, \emph{Tunable Landau-Zener transitions in a
spin-orbit-coupled Bose-Einstein condensate}, \href{http://dx.doi.org/10.1103/PhysRevA.90.013616}{
Phys. Rev. A \textbf{90}, 013616 (2014)}.

\bibitem{karina2014} K. Jimen\'{e}z-Garc\'{\i}a, L. J. LeBlanc, R. A.
Williams, M. C. Beeler, C. Qu, M. Gong, C. Zhang, and I. B. Spielman,
\emph{Tunable Spin-Orbit Coupling via Strong Driving in Ultracold-Atom Systems}, \href{http://dx.doi.org/10.1103/PhysRevLett.114.125301}%
{Phys. Rev. Lett. \textbf{114}, 125301 (2015)}.

\bibitem{Campbell2015}D. L. Campbell, R. M. Price, A. Putra, A. Vald\'es-Curiel, D. Trypogeorgos, and I. B.
Spielman, \emph{Itinerant magnetism in spin-orbit coupled Bose
gases} \href{http://arxiv.org/abs/1501.05984}{arXiv:1501.05984}.

% SOC experiments on fermions

\bibitem{Wang2012} P. Wang, Z.-Q. Yu, Z. Fu, J. Miao, L. Huang, S. Chai, H.
Zhai, and J. Zhang, \emph{Spin-Orbit Coupled Degenerate Fermi
Gases}, \href{http://dx.doi.org/10.1103/PhysRevLett.109.095301}{
Phys. Rev. Lett. \textbf{109}, 095301 (2012)}.

\bibitem{Cheuk2012} L. W. Cheuk, A. T. Sommer, Z. Hadzibabic, T. Yefsah, W.
S. Bakr, and M. W. Zwierlein, \emph{Spin-Injection Spectroscopy of
a Spin-Orbit Coupled Fermi Gas}, \href{http://dx.doi.org/10.1103/PhysRevLett.109.095302}%
{Phys. Rev. Lett. \textbf{109}, 095302 (2012)}.

\bibitem{Williams2013} R. A. Williams, M. C. Beeler, L. J. LeBlanc, K. Jim%
\'{e}nez-Garc\'{\i}a, and I. B. Spielman, \emph{Raman-Induced
Interactions in a Single-Component Fermi Gas Near an $s$-Wave
Feshbach Resonance}, \href{http://dx.doi.org/10.1103/PhysRevLett.111.095301}%
{Phys. Rev. Lett. \textbf{111}, 095301 (2013).}

\bibitem{Fu2014}Z. Fu, L. Huang, Z. Meng, P. Wang, L. Zhang, S. Zhang, H. Zhai, P. Zhang, and J.
Zhang, \emph{Production of Feshbach molecules induced by
spin-orbit coupling in Fermi gases},
\href{http://dx.doi.org/10.1038/nphys2824}{Nat. Phys. \textbf{10},
110 (2014)}.

% SOC reviews
\bibitem{Galitski2013} V. Galitski and I. B. Spielman, \emph{Spin-orbit coupling in quantum gases}, \href{http://dx.doi.org/10.1038/nature11841}%
{Nature \textbf{495}, 49 (2013)}.

\bibitem{Zhou2013} X. Zhou, Y. Li, Z. Cai, C. Wu, \emph{Unconventional states of bosons with the synthetic spin-orbit coupling}, \href{http://dx.doi.org/10.1088/0953-4075/46/13/134001}%
{J. Phys. B: At. Mol. Opt. Phys. \textbf{46}, 134001 (2013)}.

\bibitem{Zhai2015}H. Zhai, \emph{Degenerate quantum gases with spin-orbit coupling: a
review},
\href{http://dx.doi.org/10.1088/0034-4885/78/2/026001}{Rep. Prog.
Phys. \textbf{78}, 026001 (2015)}.

% SOC theory on bosons

\bibitem{Wang2010} C. Wang, C. Gao, C.-M. Jian, and H. Zhai, \emph{Spin-Orbit Coupled Spinor Bose-Einstein Condensates}, \href{http://dx.doi.org/10.1103/PhysRevLett.105.160403}%
{Phys. Rev. Lett. \textbf{105}, 160403 (2010)}.

\bibitem{Wu2011} C. Wu, I. Mondragon-Shem, and X.-F. Zhou, \emph{Unconventional Bose-Einstein Condensations from Spin-Orbit Coupling}, \href{http://dx.doi.org/10.1088/0256-307X/28/9/097102}%
{Chin. Phys. Lett. \textbf{28}, 097102 (2011)}.

\bibitem{Ho2011} T.-L. Ho and S. Zhang, \emph{Bose-Einstein Condensates with Spin-Orbit Interaction}, \href{http://dx.doi.org/10.1103/PhysRevLett.107.150403}%
{Phys. Rev. Lett. \textbf{107}, 150403 (2011)}.

\bibitem{Zhang2012} Y. Zhang, L. Mao, and C. Zhang, \emph{Mean-Field Dynamics of Spin-Orbit Coupled Bose-Einstein Condensates}, \href{http://dx.doi.org/10.1103/PhysRevLett.108.035302}%
{Phys. Rev. Lett. \textbf{\ 108}, 035302 (2012)}.

\bibitem{Hu2012} H. Hu, B. Ramachandhran, H. Pu, and X.-J. Liu, \emph{Spin-Orbit Coupled Weakly Interacting Bose-Einstein Condensates in Harmonic Traps}, \href{http://dx.doi.org/10.1103/PhysRevLett.108.010402}%
{Phys. Rev. Lett. \textbf{108}, 010402 (2012)}.

\bibitem{Ozawa2012} T. Ozawa and G. Baym, \emph{Stability of Ultracold Atomic Bose Condensates with Rashba Spin-Orbit Coupling against Quantum and Thermal Fluctuations}, \href{http://dx.doi.org/10.1103/PhysRevLett.109.025301}%
{Phys. Rev. Lett. \textbf{109}, 025301 (2012)}.

\bibitem{Li2012} Y. Li, L. P. Pitaevskii, and S. Stringari, \emph{Quantum Tricriticality and Phase Transitions in Spin-Orbit Coupled Bose-Einstein Condensates}, \href{http://dx.doi.org/10.1103/PhysRevLett.108.225301}%
{Phys. Rev. Lett. \textbf{108}, 225301 (2012)}.

\bibitem{Xu2013} Y. Xu, Y. Zhang, and B.Wu, \emph{Bright solitons in spin-orbit-coupled Bose-Einstein condensates}, \href{http://dx.doi.org/10.1103/PhysRevA.87.013614}%
{Phys. Rev. A \textbf{87}, 013614 (2013)}.

\bibitem{Zhang2013}Y. Zhang and C. Zhang, \emph{Bose-Einstein condensates in spin-orbit-coupled optical lattices: Flat bands and
superfluidity},
\href{http://dx.doi.org/10.1103/PhysRevA.87.023611}{Phys. Rev. A
\textbf{87}, 023611 (2013)}.

\bibitem{Wei2013} R. Wei and E. J. Mueller, \emph{Magnetic-field dependence of Raman coupling in alkali-metal atoms}, \href{http://dx.doi.org/10.1103/PhysRevA.87.042514}%
{Phys. Rev. A \textbf{87}, 042514 (2013)}.

\bibitem{Fetter2014} A. L. Fetter, \emph{Vortex dynamics in spin-orbit-coupled Bose-Einstein
condensates},
\href{http://dx.doi.org/10.1103/PhysRevA.89.023629}{Phys. Rev. A
\textbf{89}, 023629 (2014)}.

% SOC theory on fermions

\bibitem{Gong2011} M. Gong, S. Tewari, and C. Zhang, \emph{BCS-BEC Crossover and Topological Phase Transition in 3D Spin-Orbit Coupled Degenerate Fermi Gases}, \href{http://dx.doi.org/10.1103/PhysRevLett.107.195303}%
{Phys. Rev. Lett. \textbf{107}, 195303 (2011)}.

\bibitem{Hu2011} H. Hu, L. Jiang, X.-J. Liu, and H. Pu, \emph{Probing Anisotropic Superfluidity in Atomic Fermi Gases with Rashba Spin-Orbit Coupling}, \href{http://dx.doi.org/10.1103/PhysRevLett.107.195304}%
{Phys. Rev. Lett. \textbf{107}, 195304 (2011)}.

\bibitem{Yu2011} Z.-Q. Yu and H. Zhai, \emph{Spin-Orbit Coupled Fermi Gases across a Feshbach Resonance}, \href{http://dx.doi.org/10.1103/PhysRevLett.107.195305}%
{Phys. Rev. Lett. \textbf{107}, 195305 (2011)}.

\bibitem{Qu13}C. Qu, Z. Zheng, M. Gong, Y. Xu, Li Mao, X. Zou, G. Guo, and C Zhang, \emph{Topological superfluids with finite-momentum pairing and Majorana fermions}, \href{http://dx.doi.org/10.1038/ncomms3710}{Nat. Commun. \textbf{4}, 2710 (2013)}.
\bibitem{Zhang13}W. Zhang and W. Yi, \emph{Topological Fulde-Ferrell-Larkin-Ovchinnikov states in spin-orbit-coupled Fermi gases}, \href{http://dx.doi.org/10.1038/ncomms3711}{Nat. Commun. \textbf{4}, 2711 (2013)}.
\bibitem{Chen13}C. Chen, \emph{Inhomogeneous Topological Superfluidity in One-Dimensional Spin-Orbit-Coupled Fermi Gases}, \href{http://dx.doi.org/10.1103/PhysRevLett.111.235302}{Phys. Rev. Lett. \textbf{111}, 235302 (2013)}.

\bibitem{Xu2014}Y. Xu, C. Qu, M. Gong, and C. Zhang, \emph{Competing superfluid orders in spin-orbit-coupled fermionic cold-atom optical
lattices},
\href{http://dx.doi.org/10.1103/PhysRevA.89.013607}{Phys. Rev. A
\textbf{89}, 013607 (2014)}.

\bibitem{Lin14} F. Lin, C. Zhang, and V. W. Scarola, \emph{Emergent Kinetics and Fractionalized Charge in 1D Spin-Orbit Coupled Flatband Optical
Lattices},
\href{http://dx.doi.org/10.1103/PhysRevLett.112.110404}{Phys. Rev.
Lett. \textbf{112}, 110404 (2014)}.

\bibitem{Xu2014a}Y. Xu, L. Mao, B. Wu, and C. Zhang, \emph{Dark Solitons with Majorana Fermions in Spin-Orbit-Coupled Fermi
Gases},
\href{http://dx.doi.org/10.1103/PhysRevLett.113.130404}{Phys. Rev.
Lett. \textbf{113}, 130404 (2014)}.

\bibitem{Jiang2014}L. Jiang, E. Tiesinga, X.-J. Liu, H. Hu, and H.
Pu, \emph{Spin-orbit-coupled topological Fulde-Ferrell states of
fermions in a harmonic trap},
\href{http://dx.doi.org/10.1103/PhysRevA.90.053606}{Phys. Rev. A
\textbf{90}, 053606 (2014)}.

\bibitem{Xu2014b}Y. Xu and C. Zhang, \emph{Berezinskii-Kosterlitz-Thouless Phase Transition in 2D Spin-Orbit Coupled Fulde-Ferrell
Superfluids},
\href{http://dx.doi.org/10.1103/PhysRevLett.114.110401}{Phys. Rev.
Lett. \textbf{114}, 110401 (2015)}.

% Atom-light interaction

\bibitem{Spielman2009} I. B. Spielman, \emph{Raman processes and effective gauge potentials}, \href{http://dx.doi.org/10.1103/PhysRevA.79.063613}%
{Phys. Rev. A \textbf{79}, 063613 (2009)}.

\bibitem{Dalibard2011} J. Dalibard, F. Gerbier, G. Juzeli\={u}nas, and P.
\"{O}hberg, \emph{Colloquium: Artificial gauge potentials for
neutral atoms},
\href{http://dx.doi.org/10.1103/RevModPhys.83.1523}{Rev. Mod.
Phys. \textbf{83}, 1523 (2011)}.

\bibitem{Goldman2013} N. Goldman, G. Juzeli\={u}nas, P. \"{O}hberg, I. B.
Spielman, \emph{Light-induced gauge fields for ultracold atoms},
\href{http://dx.doi.org/10.1088/0034-4885/77/12/126401}{Rep. Prog.
Phys. \textbf{77}, 126401 (2014)}.

% SOAMC

\bibitem{Sun2014} K. Sun, C. Qu, and C. Zhang, \emph{Spin--orbital angular momentum coupled Bose-Einstein condensates}, \href{http://arxiv.org/abs/1411.1737}
{arXiv:1411.1737}.

\bibitem{Hu} Y.-X. Hu, C. Miniatura, and B. Gr\'{e}maud, \emph{Half-skyrmion and meron pair in spinor condensates}, \href{http://arxiv.org/abs/1410.8634}
{arXiv:1410.8634}.

\bibitem{Pu} M. Demarco, and H. Pu, \emph{Angular spin-orbit coupling in cold atoms}, \href{http://dx.doi.org/10.1103/PhysRevA.91.033630}{
Phys. Rev. A \textbf{91}, 033630 (2015)}.

\bibitem{Marzlin1997} K.-P. Marzlin, W. Zhang, and E. M. Wright, \emph{Vortex Coupler for Atomic Bose-Einstein Condensates}, \href{http://dx.doi.org/10.1103/PhysRevLett.79.4728}
{Phys. Rev. Lett. \textbf{79}, 4728 (1997)}.

\bibitem{Juzeliunas2004} G. Juzeli\={u}nas and P. \"{O}hberg, \emph{Slow Light in Degenerate Fermi Gases}, \href{http://dx.doi.org/10.1103/PhysRevLett.93.033602}
{Phys. Rev. Lett. \textbf{93}, 033602 (2004)}.

\bibitem{Cooper2010} N. R. Cooper and Z. Hadzibabic, \emph{Measuring the Superfluid Fraction of an Ultracold Atomic Gas}, \href{http://dx.doi.org/10.1103/PhysRevLett.104.030401}
{Phys. Rev. Lett. \textbf{104}, 030401 (2010)}.

\bibitem{Ramanathan2011} A. Ramanathan, K. C. Wright, S. R. Muniz, M. Zelan,
W. T. Hill, C. J. Lobb, K. Helmerson, W. D. Phillips, and G. K. Campbell,
\emph{Superflow in a Toroidal Bose-Einstein Condensate: An Atom Circuit with a Tunable Weak Link},
\href{http://dx.doi.org/10.1103/PhysRevLett.106.130401}{Phys. Rev. Lett.
\textbf{106}, 130401 (2011)}.

\bibitem{Ho1996} T.-L. Ho and V. B. Shenoy, \emph{Binary Mixtures of Bose Condensates of Alkali Atoms}, \href{http://dx.doi.org/10.1103/PhysRevLett.77.3276}
{Phys. Rev. Lett. \textbf{77}, 3276 (1996).}

\bibitem{Ueda} K. Kasamatsu, M. Tsubota, and M. Ueda, \emph{Vortex Molecules in Coherently Coupled Two-Component Bose-Einstein Condensates}, \href{http://dx.doi.org/10.1103/PhysRevLett.93.250406}%
{Phys. Rev. Lett. \textbf{93}, 250406 (2004).}

\bibitem{Andersen2006} M. F. Andersen, C. Ryu, P. Clad\'{e}, V. Natarajan,
A. Vaziri, K. Helmerson, and W. D. Phillips, \emph{Quantized
Rotation of Atoms from Photons with Orbital Angular Momentum},
\href{http://dx.doi.org/10.1103/PhysRevLett.97.170406} {Phys. Rev.
Lett. \textbf{97}, 170406 (2006)}.

\bibitem{Ryu2007} C. Ryu, M. F. Andersen, P. Clad\'{e}, V. Natarajan, K.
Helmerson, and W. D. Phillips, \emph{Observation of Persistent
Flow of a Bose-Einstein Condensate in a Toroidal Trap},
\href{http://dx.doi.org/10.1103/PhysRevLett.99.260401} {Phys. Rev.
Lett. \textbf{99}, 260401 (2007)}.

\bibitem{Leslie2009} L. S. Leslie, A. Hansen, K. C. Wright, B. M. Deutsch,
and N. P. Bigelow, \emph{Creation and Detection of Skyrmions in a
Bose-Einstein Condensate}, \href{http://dx.doi.org/10.1103/PhysRevLett.103.250401}{
Phys. Rev. Lett. \textbf{103}, 250401 (2009)}

\bibitem{Gullo2010} N. Lo Gullo, S. McEndoo, T. Busch, and M. Paternostro, \emph{Vortex entanglement in Bose-Einstein
condensates coupled to Laguerre-Gauss beams},
\href{http://dx.doi.org/10.1103/PhysRevA.81.053625}{Phys. Rev. A
\textbf{81}, 053625 (2010)}.

\bibitem{Beattle2013} S. Beattie, S. Moulder, R. J. Fletcher, and Z.
Hadzibabic, \emph{Persistent Currents in Spinor Condensates}, \href{http://dx.doi.org/10.1103/PhysRevLett.110.025301}{Phys.
Rev. Lett. \textbf{110}, 025301 (2013)}.

\bibitem{Moulder2012} S. Moulder, S. Beattie, R. P. Smith, N. Tammuz, and Z.
Hadzibabic, \emph{Quantized supercurrent decay in an annular Bose-Einstein condensate}, \href{http://dx.doi.org/10.1103/PhysRevA.86.013629}{Phys. Rev. A
\textbf{86}, 013629 (2012)}.

\bibitem{DAMOP} P.-P. Huang, C.-A. Chen, H.-J. Wei, C.-Y. Yu, J.-B. Wang,
and Y.-J. Lin, \emph{Towards generating synthetic gauge potentials
for a Bose-Einstein condensate in a toroidal trap}, 2015 DAMOP
(forthcoming) abstract,
\href{http://meetings.aps.org/Meeting/DAMOP15/Session/K1.53}{http://meetings.aps.org/Meeting/DAMOP15/Session/K1.53}.

\end{thebibliography}
\end{document}